\documentclass[11pt]{article}
\usepackage[utf8]{inputenc}
\usepackage{authblk}
\usepackage{amsmath}
\usepackage{amssymb}
\usepackage{graphicx}
\usepackage{xcolor}
\usepackage{setspace}
\usepackage[margin=1in]{geometry}
\usepackage{hyperref}

\title{Fluidity and morphological stability of an amorphous thin film with radiation-induced defect kinetics}
\author[1]{Tyler P. Evans \footnote{corresponding author, evans.tyler@utah.edu, https://orcid.org/0000-0001-7812-2479 } }
\author[2]{Eden Heyen}
\affil[1]{Department of Mathematics, University of Utah, Salt Lake City, UT 84112, United States of America}
\affil[2]{Department of Mathematics, Colorado State University, Ft. Collins, CO 80523 United States of America}

\begin{document}

\maketitle

\begin{abstract}
It is common to model ion-irradiated amorphous thin films as if they were highly viscous fluids. In such models, one is frequently concerned with the \textit{ion-enhanced fluidity}, a measure of the ability of the free interface to relax surface energy. Motivated by usual fluid dynamics problems, the ion-enhanced fluidity is near-universally treated as a constant throughout the amorphous layer. However, for an irradiated thin film, the fluidity is ultimately caused by radiation-induced defect kinetics within the film, leading to regions with greater or lesser fluidity, and sensitive dependence on ion energy, species, flux, irradiation angle, temperature, and other experimental parameters. Here, we develop and analyze a model of radiation-induced defect kinetics coupled to the continuum equations of a viscous thin film. Using realistic parameter values, we show that defect kinetics can meaningfully alter theoretical predictions of surface relaxation and ion-enhanced fluidity. Implications for aligning theoretical and experimental work, especially for surface nano-patterning of silicon induced by low-energy argon irradiation, are discussed.
\end{abstract}

\tableofcontents

\section{Introduction}
Broad-beam, low-energy ion bombardment of group IV semiconductors, such as silicon and germanium, leads to the destruction of the crystal lattice and the formation of an amorphous layer whose thickness is typically on the order of a few nanometers \cite{umbach-etal-PRL-2001}. Since the late 1990s, it has been proposed that stress accumulation and relaxation in this thin, amorphous layer are important for determining the dynamics of semiconductor surfaces under irradiation \cite{rudy-smirnov-NIMB-1999,umbach-etal-PRL-2001,chan-chason-JAP-2007,munoz-garcia-etal-MSER-2014,NorrisAziz_predictivemodel,cuerno-kim-JAP-2020-perspective}. Many contemporary models of the dynamics of these amorphous thin films posit that they behave as fluids with extremely high viscosity $\eta$ and aim to understand them using techniques from fluid mechanics, and especially hydrodynamic stability theory \cite{cuerno-etal-NIMB-2011,castro-cuerno-ASS-2012,castro-etal-PRB-2012,norris-PRB-2012-viscoelastic-normal,norris-PRB-2012-linear-viscous,chandrasekhar-book-2013,moreno-barrado-etal-PRB-2015,norris-etal-SREP-2017,Swenson_2018,munoz-garcia-etal-PRB-2019,NorrisAziz_predictivemodel,evans-norris-JPCM-2022,evans-norris-JPCM-2023,evans-norris-JEM-2024}, or pattern formation theory \cite{cross-hohenberg-RoMP-1993,cross-greenside-book}.

Here, caution is needed: the analogy between classical fluids and irradiated thin films is imperfect. Stress accumulation and relaxation in the irradiated thin film are largely determined by implanted ions and the kinetics of \textit{flow defects} (Frenkel pairs) created during irradiation \cite{kinchin-pease-1955,davis-TSF-1993-simple-compressive-stress,chan-chason-JAP-2007,chan-chason-JVSTA-2008,ishii-etal-JMR-2014}. These flow defects produce an \textit{ion-enhanced fluidity}, $\eta^{-1}$, which allows surface energy relaxation \cite{orchard-ASR-1962}. Exact values of $\eta^{-1}$ for experimental systems of interest have been subject to much speculation due to the central importance that $\eta^{-1}$ holds in modeling surface evolution in viscous models \cite{norris-etal-NCOMM-2011,ishii-etal-JMR-2014,hofsass-bobes-zhang-JAP-2016,norris-etal-SREP-2017,NorrisAziz_predictivemodel}. For lack of a comprehensive, first-principles model, it has been commonly used as a \textit{de facto} fit parameter \cite{norris-etal-NCOMM-2011,hofsass-APA-2015,moreno-barrado-etal-PRB-2015,NorrisAziz_predictivemodel}, and both linear and nonlinear scalings of $\eta^{-1}$ with ion flux and energy have been discussed \cite{norris-etal-NCOMM-2011,ishii-etal-JMR-2014,hofsass-bobes-zhang-JAP-2016}. When the angle dependence of $\eta^{-1}$ has been considered, it has been assumed to follow $\cos(\theta)$ \cite{norris-etal-SREP-2017,NorrisAziz_predictivemodel}.

In modeling surface evolution for irradiated semiconductors, a near-universal assumption is that $\eta^{-1}$ is constant throughout the film, as is typically the case in traditional fluid mechanics problems, leading to surface energy relaxation according to the analysis of \cite{orchard-ASR-1962}. However, one does not generally expect that the distribution of defects --- and therefore the ion-enhanced fluidity --- should be constant throughout an irradiated film; see, e.g., \cite{kalyanasundaram-AM-2006,chan-chason-JVSTA-2008,moreno-barrado-etal-PRB-2015}. Adding to the danger of this common simplification, it has now been seen in several other contexts that spatial inhomogeneity of model parameters can introduce significant differences in model predictions of irradiated surface evolution: spatially inhomogenous energy distribution produces the well-known Bradley-Harper instability \cite{bradley-harper-JVST-1988,bradley-PRB-2011b}; local differences in density produce a stabilizing effect \cite{Swenson_2018,evans-norris-JPCM-2023}; and tying local density changes directly to ion implantation can greatly strengthen or weaken that stabilizing effect \cite{evans-norris-arxiv-2025b}.

To understand how the defect kinetics determine the amorphous layer's ion-enhanced fluidity and induce a complex dependence of the fluidity on ion flux, energy, film temperature, and the distribution of defects within the amorphous layer, we study a generalization of a model due to \cite{chan-chason-JVSTA-2008}. This model was originally used to simulate defect kinetics resulting from normal-incidence irradiation of a flat substrate and explained residual stresses and thermal effects in a series of wafer-curvature experiments on irradiated copper, which does not become amorphous. Here, we adapt the defect kinetics model from \cite{chan-chason-JVSTA-2008} to the case of off-normal incidence and couple it to a continuum model of a viscous fluid with a free surface. We use the resulting model to address two related questions: 
\begin{itemize}
    \item When defect kinetics in the style of \cite{chan-chason-JVSTA-2008,ishii-etal-JMR-2014} are explicitly included in free-surface continuum models of an amorphous thin film, do any new phenomena emerge --- particularly, new morphological instabilities, or qualitative changes in viscous surface relaxation?
    \item Can a continuum model of the type studied here lead to quantitatively-reasonable predictions in how the fluidity should change with flux, ion energy, irradiation angle, and other model parameters?
\end{itemize}
Throughout the history of theoretically modeling ion-induced pattern formation, multiple surface relaxation mechanisms have been proposed, used, and later refuted. The initial Bradley-Harper model used surface diffusion as a surface relaxation mechanism \cite{bradley-harper-JVST-1988}, and later removed following \cite{umbach-etal-PRL-2001}. An effective surface diffusion mechanism was proposed \cite{makeev-barabasi-APL-1997,makeev-etal-NIMB-2002}, and subsequently refuted \cite{bradley-PRB-2011b,NorrisAziz_predictivemodel}, before broad agreement was reached that viscous surface relaxation was a better explanation \cite{umbach-etal-PRL-2001,norris-etal-SREP-2017,NorrisAziz_predictivemodel,myint-ludwig-etal-PRB-2021-Ar-bombardment,myint-ludwig-PRB-2021-Kr-bombardment}. However, given the difficulty of obtaining a single, unified model for ion-induced nano-patterning \cite{NorrisAziz_predictivemodel,cuerno-kim-JAP-2020-perspective}, it is natural to revisit prevailing modeling assumptions in search of important physics that those assumptions may have obscured; a central question is how far the ``viscous" characterization of the thin film may be pushed.

The paper is organized as follows. First, we discuss the continuum model studied in the present work, which is motivated by background on both viscous thin films and a pre-existing model for radiation-induced defect kinetics. Second, we state the mathematical problem of interest: the steady-state equations, the derivation of equations for the growth of a single Fourier mode, a simplifying limit, and some details of numerical implementation. Finally, we consider the results from the model. We (i) compare linear-regime surface relaxation for thin films of constant fluidity \cite{orchard-ASR-1962} with thin films of spatially-varying fluidity resulting from defect kinetics; (ii) comment on the use of an approximately $\cos(\theta)$ angle-dependence of the mean fluidity, $\eta^{-1}(\theta)$ \cite{norris-etal-SREP-2017}; and (iii) explore of flux and energy dependence of $\eta^{-1}(\theta)$, with an eye towards validating commonly used analytical approximations which do not explicitly consider the defect kinetics \cite{norris-etal-NCOMM-2011,norris-etal-SREP-2017,NorrisAziz_predictivemodel}.

\section{Model}
To describe the dynamics of the first few nanometers of the irradiated surface, we follow the approach found in the substantial literature on viscous flow models of ion-induced nanopatterning \cite{rudy-smirnov-NIMB-1999,umbach-etal-PRL-2001,cuerno-etal-NIMB-2011,castro-cuerno-ASS-2012,castro-etal-PRB-2012,norris-PRB-2012-viscoelastic-normal,norris-PRB-2012-linear-viscous,moreno-barrado-etal-PRB-2015,Swenson_2018,munoz-garcia-etal-PRB-2019,evans-norris-JPCM-2022,evans-norris-JPCM-2023,evans-norris-JEM-2024,evans-norris-arxiv-2025} where this heavily damaged region is treated as a fluid of extremely high viscosity $\eta$. It is to this viscous-type model that we connect a separate model of defect kinetics \cite{chan-chason-JVSTA-2008}.
\paragraph{Conservation laws.} Assuming an isothermal viscous fluid, the constitutive law
\begin{equation}
\textbf{T} = - p(x,z,t) \textbf{I} + 2\eta(x,z,t) \dot{\textbf{E}} \label{stresstensor}
\end{equation}
is applied, with $\textbf{T}$ the Cauchy stress tensor, $\mathbf{I}$ the identity tensor, $p$ the hydrostatic pressure, and $\dot{\textbf{E}} = \frac{1}{2}\big(\nabla \vec{v} + \nabla \vec{v}^T \big)$ the usual strain rate tensor. Here, $\eta(x,z,t)$ is the \textit{local} viscosity, an important departure from existing work. Conservation of mass and momentum, respectively, are applied in differential form as 
\begin{equation}
\begin{gathered}
\frac{\partial \rho}{\partial t} + \nabla \cdot \big(\rho \vec{v}\big) = 0 \\
\nabla \cdot \textbf{T} = \vec{0}.
\end{gathered}
\end{equation}
where $\rho=\rho(x,z,t)$ is the local mass density and $\vec{v}$ is the local Eulerian velocity field. Above, conservation of momentum is written in the limit of vanishingly-small Reynolds number (Re $\to 0$), as is standard for these systems; see, e.g., \cite{cuerno-etal-NIMB-2011,castro-cuerno-ASS-2012,castro-etal-PRB-2012,norris-PRB-2012-viscoelastic-normal,norris-PRB-2012-linear-viscous,moreno-barrado-etal-PRB-2015}. With Equation \ref{stresstensor}, momentum conservation specializes to 
\begin{equation}
-\nabla p + 2\nabla \cdot \big(\eta \dot{\textbf{E}} \big) = \vec{0}.
\end{equation}
These standard continuum equations for a viscous fluid are then coupled to the defect kinetics in the amorphous bulk. We model the defect kinetics as a generalization of \cite{chan-chason-JVSTA-2008}, where attention was restricted to normal-incidence irradiation and viscosity was not considered due to the experimental system of interest. The concentration fields of vacancies $C_V$ and interstitials $C_I$, which have units nm$^{-3}$, are governed by the reaction-diffusion equations
\begin{equation}
\begin{gathered}
\frac{\partial C_{V} }{\partial t} = fY_VP(x,z;\theta) + D_{V} \nabla^2 C_V - \frac{D_{V}C_{V}V_{V}}{K_B T}\frac{1}{3}\nabla^2\text{tr}\big(\textbf{T}\big) - 4\pi \lambda C_I C_V (D_I + D_V) \\
\frac{\partial C_{I} }{\partial t} = fY_IP(x,z;\theta) + D_{I} \nabla^2 C_I - \frac{D_{I}C_{I}V_{I}}{K_B T} \frac{1}{3}\nabla^2\text{tr}\big(\textbf{T}\big) - 4\pi \lambda C_I C_V (D_I + D_V) \label{defect-kinetics}
\end{gathered}
\end{equation}
with $f$ the nominal flux of ions, $Y_V$ and $Y_I$ local defect production rates for vacancies and interstitials, respectively. Defect production varies spatially throughout the amorphous layer according to the distribution of deposited energy, $P(x,z;\theta)$ \cite{kinchin-pease-1955,ziegler-biersack-littmark-1985-SRIM,srim-2000.40}. Here, $\theta$ is the angle of irradiation against a macroscopically flat surface, with $\theta=0^{\circ}$ normal-incidence irradiation, and $\theta \to 90^{\circ}$ grazing-incidence irradiation. $D_I$ and $D_V$ are diffusivities of interstitials and vacancies, respectively. $V_I$ and $V_V$ are the relaxation volumes of interstitials and vacancies \cite{chan-chason-JVSTA-2008}. $K_B$ is the Boltzmann constant and $T$ is the temperature in the thin film, which can be estimated theoretically \cite{parry-JVST-1976,albaugh-et-al-JVSTB-1990}, and is assumed constant in space and time. $\lambda$ is the mean distance over which an interstitial and a vacancy can recombine.

In (\ref{defect-kinetics}), the terms on the right of each equation represent (i) the creation of new defects due to the influence of the ion beam, (ii) diffusion of defects driven by the concentration gradient, (iii) diffusion of defects driven by the stress gradient, and (iv) annihilation of defects by recombination of vacancy-interstitial pairs within interaction range $\lambda$.

\paragraph{Equations of state and fluidity.} For closure, we require an equation of state. Somewhat generalizing \cite{Swenson_2018,evans-norris-JPCM-2022,evans-norris-JPCM-2023}, we apply a simple ``quasi-incompressibility" equation of state, 
\begin{equation}
    \rho(x,z,t) = \frac{m_0}{V_0 + C_IV_I + C_VV_V},
\end{equation}
where $m_0$ is atomic mass and $V_0$ is atomic volume. This says that density changes can only result from local defects, and are not explicitly due to pressure. Despite its simplicity, similar equations of state have been successfully used elsewhere \cite{Swenson_2018,evans-norris-JPCM-2022,evans-norris-JPCM-2023,kozlovskiy-et-al-JCS-2023,moldavayeva-et-al-ESMaM-2024,garcia-et-al-NIMB-2024,evans-norris-JEM-2024,evans-norris-arxiv-2025} as an approximation. Note that identifying relative volumization $\Delta = C_I\frac{V_I}{V_0} + C_V\frac{V_V}{V_0}$ and basic density $\rho^* = \frac{m_0}{V_0}$ recovers the form
\begin{equation}
    \rho(x,z,t) = \frac{\rho^*}{1 + \Delta}
\end{equation}
which has been established elsewhere \cite{Swenson_2018,evans-norris-JPCM-2023}. As a description of the ion-enhanced fluidity, we suppose
\begin{equation}
    \frac{1}{\eta} = \frac{1}{\eta^*} + \bigg(\frac{C_I}{\eta_I} + \frac{C_V}{\eta_V} \bigg),
\end{equation}
or, equivalently, the viscosity $\eta = \big[\frac{1}{\eta^*} + \big(\frac{C_I}{\eta_I} + \frac{C_V}{\eta_V}\big)\big]^{-1}$. This makes the assumption, which has been used elsewhere \cite{norris-etal-NCOMM-2011,ishii-etal-JMR-2014}, that the local ion-enhanced fluidity is linear in the number of flow defects, with $\eta_I^{-1}$ the contribution to fluidity by interstitials and $\eta_V^{-1}$ the contribution to fluidity by vacancies. The parameter $1/\eta^{*}$ is a baseline fluidity in the absence of any defects, which is assumed to be very small.

\paragraph{Modeling defect production.} In the governing equations of the defect fields, we take the flux-diluted energy deposition \cite{sigmund-PR-1969,sigmund-JMS-1973,bradley-harper-JVST-1988,chan-chason-JAP-2007,bradley-PRB-2011b,NorrisAziz_predictivemodel} through a free surface $h(X,t)$,
\begin{equation}
	P(x,z,t) = \int_{-\infty}^{\infty}\frac{\cos(\theta) + h_X\sin(\theta)}{\sqrt{1+h_X^2}} E(x,z;X,h(X,t))dX \label{defects-bulk}
\end{equation}
where
\begin{equation}
    E(x,z; X, h(X,t)) = \frac{E_0}{2\pi\alpha \beta}e^{\big(-\frac{[(x-X)\sin(\theta) - (z-h(X,t))\cos(\theta)-a]^2}{2\alpha^2} - \frac{[(x-X)\cos(\theta) + (z-h(X,t))\sin(\theta)]^2}{2\beta^2} \big)},
\end{equation}
is the energy deposited at a given location in the film, which results in the creation of Frenkel pairs. Elsewhere, it has been shown that $E(x,z; X, h(X,t))$ is reasonably well-approximated as a bivariate Gaussian ellipsoid in downbeam-crossbeam coordinates \cite{sigmund-PR-1969,sigmund-JMS-1973,bradley-harper-JVST-1988,bradley-PRB-2011b,hossain-etal-JAP-2012,hobler-etal-PRB-2016,chan-chason-JAP-2007}. Here, $a$ is the downbeam depth where energy deposition is greatest, corresponding to the downbeam depth that ions penetrate before initiating the collision cascade, $\alpha$ is the downbeam standard deviation, and $\beta$ is the crossbeam standard deviation.

\paragraph{Boundary conditions.} As the surface is irradiated, material is sputtered away from the free interface \cite{sigmund-PR-1969}, and newly-exposed crystalline substrate can be amorphized. Hence the present model describes a thin film translating in the direction of the ion beam, as in \cite{Swenson_2018,evans-norris-JPCM-2022,evans-norris-JPCM-2023,evans-norris-JEM-2024}, where a translating reference frame is introduced. The ongoing erosion of the free interface leads to the modified kinematic equation \cite{Swenson_2018}, (\ref{freeBCs}b), below, and an apparent non-zero velocity at the amorphous-crystalline interface, (\ref{lowerBCs}a) \cite{Swenson_2018,evans-norris-JPCM-2022,evans-norris-JPCM-2023,evans-norris-JEM-2024}. At the free surface $z=h(x,t)$,
\begin{equation}
\begin{gathered}
    \mathbf{T} \cdot \hat{n} = \gamma \kappa \mathbf{\hat{n}} \\
    v_I = \vec{v}\cdot \mathbf{\hat{n}} - V(\theta)\frac{\rho^*}{\rho} \label{freeBCs}
\end{gathered}
\end{equation}
where $\gamma$ is surface energy, and $\kappa$ is the mean curvature; in one dimension, $\kappa= \frac{h_{xx}}{(1 + h_{x}^2)^{3/2}}$. $\mathbf{\hat{n}}$ is the outward-normal unit vector from the free surface, $v_I$ is the outward-normal interfacial speed (i.e., $\frac{h_t}{\sqrt{1+h_x^2}}$ in one dimension), and $V(\theta)$ is the speed of downward translation (in laboratory coordinates) of the film due to erosion \cite{sigmund-PR-1969,sigmund-JMS-1973,bradley-harper-JVST-1988,bradley-PRB-2011b}. The first equation above is stress-balance at the interface and the second is the kinematic equation modified to account for ongoing sputtering \cite{Swenson_2018}. At the lower interface $z=g(x,t)$,
\begin{equation}
\begin{gathered}
    \vec{v} = V(\theta)\hat{k} \\
    \Delta = 
    \frac{\partial \Delta}{\partial \hat{n}} =
    \frac{\partial^2 \Delta}{\partial \hat{n}^2} = 0, \label{lowerBCs}
\end{gathered}
\end{equation}
where $\Delta = C_I\frac{V_I}{V_0} + C_V\frac{V_V}{V_0}$, as above, and $\hat{k}$ is a unit vector along the $z$ axis. These boundary conditions for $C_I$ and $C_V$ are chosen to describe a sharp interface between the amorphous and crystalline phases. In principle, more sophisticated boundary conditions could be chosen, but we regard these as sufficient for an initial study. The relationship between $g(x,t)$ and $h(x,t)$ is characterized by the interface relation provided by \cite{evans-norris-JEM-2024}, where $g(x,t) = h(x-x_0(\theta),t) - h_0(\theta)$ and
\begin{equation}
\begin{gathered}
    h_0(\theta) = a\cos(\theta) + 2\sqrt{\alpha^2\cos^2(\theta) + \beta^2\sin^2(\theta)} \\
    x_0(\theta) = a\sin(\theta) + \frac{2(\alpha^2-\beta^2)\sin(\theta)\cos(\theta) }{\sqrt{\alpha^2\cos^2(\theta) + \beta^2\sin^2(\theta)}}. \label{interfaces}
\end{gathered}
\end{equation}
In order to focus on the influence of the defect kinetics, the effect of phase-change at the amorphous-crystalline interface \cite{evans-norris-JEM-2024} is ignored.

\paragraph{Discussion of model.} Some comments on this model are in order. While embracing the complexity of including defect kinetics in a model of a viscous thin film, three important simplifications are still made for tractability:
\begin{itemize}
    \item We suppose that the primary effect of implanted ions is to produce Frenkel pairs, while neglecting implantation of noble gas ions and their interactions with the amorphous layer, which were explicitly considered in \cite{chan-chason-JVSTA-2008}. This simplification is made (i) in order to leave our statement of mass conservation unchanged, (ii) to avoid the need to model bubble formation, and other effects \cite{chini-etal-PRB-2003-TEM,chan-chason-JVSTA-2008}, and (iii) to facilitate a relatively simple first study. However, by neglecting ion implantation, we sacrifice the ability to seek quantitative agreement with ion-induced stresses \cite{kalyanasundaram-AM-2006,chan-chason-JVSTA-2008}.
    \item We assume that the film that there is no meaningful thermal gradient in the film. In principle, considering the thermal gradient would mean that we would also need to include spatially-inhomogeneous diffusivities $D_I$ and $D_V$ --- a major increase in complexity on top of an already-complex model that we will defer to future work. Moreover, it has been argued elsewhere that a thin film with thickness on the order of a few nanometers should be unable to sustain a significant thermal gradient \cite{norris-PRB-2012-linear-viscous}.
    \item We do not take into account removal of defects by sputtering. We assume that the recombination of Frenkel pairs is the only means of eliminating them, as in \cite{ishii-etal-JMR-2014}. 
\end{itemize}

 Here, we have generalized \cite{chan-chason-JVSTA-2008} by using $-\frac{1}{3}\nabla^2\text{tr}(\mathbf{T})$, the divergence of the isotropic stress gradient in the film, instead of their $\frac{\partial^2\sigma}{\partial z^2}$, where $\sigma$ is the in-plane stress in their notation. This is necessary to consider off-normal incidence of irradiation.

\section{Analysis}
\paragraph{Linear stability analysis in Fourier modes.} We are interested in (i) the steady-state form of $\eta^{-1}(\theta)$ and (ii) the effect of defect dynamics on linear-regime surface relaxation, which motivates a linear stability analysis of the continuum system described above. Our linear stability analysis follows the style of \cite{cross-greenside-book}, which we briefly summarize as follows. 

For Equations (\ref{stresstensor})-(\ref{lowerBCs}), we restrict attention to the $x$ and $z$ axes, and compute the \textit{steady-state solution}. Steady-state solutions should exhibit time-invariance ($\frac{\partial}{\partial t}\to 0$) and translation-invariance along the extended axis ($\frac{\partial}{\partial x} \to 0$). Having obtained the steady-state solution, to be denoted $\vec{\Psi}_0(z) = [p_0(z),u_0(z),w_0(z),C_{V,0}(z),C_{I,0}(z)]^T$, we seek the system's linear response to vanishingly-small perturbations of the steady state via a two-term asymptotic expansion, $\vec{\Psi}(x,z,t) = \vec{\Psi}_0(z) + \epsilon \vec{\Psi}_1(x,z,t)$, where $\epsilon \approx 0$, and $\vec{\Psi}_1(x,z,t) = [p_1(x,z),u_1(x,z),w_1(x,z),C_{V,1}(x,z),C_{I,1}(x,z)]^T$ must be determined. To this end, one considers periodic boundaries on the extended $x$ axis, which motivates the use of the ansatz $e^{ikx}$ for the $x$-dependence of $\vec{\Psi}_1$. The resulting partial differential equations are separable in $t$ so that one anticipates temporal scaling $e^{\Sigma t}$. Together, this motivates the ansatz $\vec{\Psi}(x,z,t) = \vec{\Psi}_0(z) + \epsilon \vec{\bar{\Psi}}_1(z)e^{\Sigma t + ikx}$, where $\vec{\bar{\Psi}}_1(z) = [\tilde{p}_1(z),\tilde{u}_1(z),\tilde{w}_1(z),\tilde{C}_{V,1}(z),\tilde{C}_{I,1}(z)]^T$. The interfaces are similarly linearized as $h(x,t) = h_0(\theta) + \epsilon \tilde{h}_1e^{\Sigma t + ikx}$ and $g(x,t) = \epsilon \tilde{h_1}e^{\Sigma t + ik(x-x_0(\theta))}$ --- that is, one expects a corrugated lower (amorphous-crystalline) interface that is a phase-shifted and vertically-displaced copy of the upper (free) interface \cite{evans-norris-JEM-2024,evans-norris-arxiv-2025b}.

Following this, all $t$ dependence is subsumed by $\sigma$, and all $x$ dependence is subsumed by $k$. One solves the resulting system of ordinary differential equations and boundary conditions as a nonlinear eigenvalue problem for $\vec{\bar{\Psi}}_1(z)$ and the growth rate $\Sigma$ \cite{drazin-reid-book,cross-hohenberg-RoMP-1993,cross-greenside-book}. 

\paragraph{Steady-state equations in a simplifying limit.} The full steady-state equations are shown in the Appendix, but they elude analytical solution. Moreover, the equations for defect concentrations $C_I$ and $C_V$ produce a pair of coupled third-order equations that are difficult to solve, even using numerical methods. However, for some systems of interest, like amorphous silicon, the diffusion rate of vacancies, $D_V$, may be two or more orders of magnitude larger than that of interstitials, $D_I$ \cite{fahey-et-al-RoMP-1989,coffa-libertino-APL-1998}.

This suggests the use of the $\frac{D_I}{D_V} \to 0$ limit, which enables a \textit{pseudo-steady state approximation} commonly used in reaction kinetics. This simplification, described in the Appendix, produces a single nonlinear, third-order ordinary differential equation for steady-state $C_{V0}$,
\begin{equation}
\begin{gathered}
fY_V P_0(z) + D_VC_{V,0zz} + \frac{D_V}{3}\bigg(\frac{V_V}{K_BT}\bigg)C_{V,0}p_{0zz} - fY_IP_0(z) = 0, \label{steadystateCI0}
\end{gathered}
\end{equation}
when we express $p_0, \rho_0, \eta_0$ as
\begin{equation}
\begin{gathered}
    p_0 = -2V(\theta)\eta_0\frac{\rho_{0z}}{\rho_0^2}, \hspace{.25cm} 
    \rho_0 = \bigg[1 + C_{V0}\frac{V_V}{V_0} + \frac{fY_IP_0V_I}{4\pi\lambda C_{V0}D_VV_0}\bigg]^{-1}, \hspace{.25cm}
    \eta_0 = \bigg[\frac{1}{\eta^*} + \frac{C_{V0}}{\eta_V} + \frac{fY_IP_0}{4\pi\lambda C_{V0}D_V\eta_I} \bigg]^{-1}
\end{gathered}
\end{equation}
so that Equation \ref{steadystateCI0} can be written in terms of $C_{V0}(z)$, its derivatives, and assigned parameters. The resulting equation for $C_{V0}(z)$ is too long to record here, but its form is of no particular interest.

Equations \ref{lowerBCs}, again in the $D_I/D_V\to 0$ limit, lead to the following boundary conditions at the amorphous-crystalline interface $z=0$:
\begin{equation}
\begin{gathered}
    C_{V0} = \sqrt{\frac{-fY_IV_I}{4\pi\lambda D_VV_V}}\sqrt{P_0(z)} \\
    C_{V0z} = \sqrt{\frac{-fY_IV_I}{4\pi\lambda D_VV_V}}\frac{P_0'(z)}{\sqrt{P_0(z)}} \\
    C_{V0zz} = \sqrt{\frac{-fY_IV_I}{4\pi\lambda D_VV_V}}\bigg[\frac{P_0''(z)}{\sqrt{P_0(z)}} - \frac{1}{2}\frac{\big(P_0'(z)\big)^2}{P_0(z)^{3/2}}    \bigg]. \label{SSlowerBCs}
\end{gathered}
\end{equation}
Above, we have observed that $p_0, \rho_0$ and $\eta_0$ are entirely expressible in terms of $C_{V0}(z)$. In the Appendix, we also show that $w_0 = \frac{V(\theta)\rho^*}{\rho_0}$ and $u_0=0$. Hence the solution for $C_{V0}(z)$ fully determines the steady state.

The $\mathcal{O}(\epsilon)$ equations for the growth rates $\sigma$ are much more complicated, and their simplifications under the pseudo-steady state approximation are less dramatic. We defer them to the Appendix for brevity. In \cite{evans-norris-arxiv-2025b}, an approximation of ion implantation through a deformed interface, $P(x,z,t) = P_0(z) + \epsilon P_1(z)\exp(\sigma t + ikx)$, suitable to linear stability analysis for a single Fourier mode of wavenumber $k$, is computed. We use this approximation here as
\begin{equation}
    P_0(z) = \frac{\cos(\theta)}{\sqrt{2\pi(\alpha^2c^2 + \beta^2s^2)}}\exp\bigg(-\frac{(z-h_0+ac)^2}{2(\alpha^2c^2 + \beta^2s^2)} \bigg)
\end{equation}
and
\begin{equation}
\begin{gathered}
P_1(z) =  \frac{\cos(\theta)}{2\sqrt{\pi A}\alpha^3\beta^3}\exp\bigg(\frac{\tilde{B}^2(z)}{4A}-C(z) \bigg)\bigg[c_1(z)- c_2\frac{\tilde{B}(z)}{2A} \bigg]\tilde{h}_1,
\end{gathered}
\end{equation}
where
\begin{equation}
\begin{gathered}
A = \frac{\beta^2s^2 + \alpha^2c^2}{2\alpha^2\beta^2 }; \hspace{.25cm}
B(z) = \frac{(z-h_0)sc(\alpha^2-\beta^2) - a\beta^2s}{\alpha^2\beta^2}; \\
\tilde{B}(z) = \frac{(z-h_0)sc(\alpha^2-\beta^2) - a\beta^2s}{\alpha^2\beta^2} + ik; \\
C(z) = (z-h_0)^2\frac{\beta^2c^2+\alpha^2s^2}{2\alpha^2\beta^2} + \frac{a(z-h_0)c}{\alpha^2} + \frac{a^2}{2\alpha^2}; \\
c_1(z) = (z-h_0)\left( \alpha^2\sin^2(\theta) + \beta^2\cos^2(\theta)\right) + \cos(\theta)a\beta^2 + ik\alpha^2\beta^2\tan(\theta); \\
c_2 = \cos(\theta)\sin(\theta)(\alpha^2-\beta^2),
\end{gathered}
\end{equation}
and $c=\cos(\theta), s=\sin(\theta)$ throughout. The above $P_0(z)$ and $P_1(z)$ provide the local rates of energy deposition into the film. This is then converted to a point-wise rate of defect production using the Kinchin-Pease relation \cite{kinchin-pease-1955,chan-chason-JVSTA-2008}.

\paragraph{Estimating film temperature $T$.} Similar to \cite{parry-JVST-1976,albaugh-et-al-JVSTB-1990}, we estimate the temperature of the irradiated wafer using the point model
\begin{equation}
    \frac{dT}{dt} = \frac{1}{\rho_c C_vh_0}\bigg(fE_{ion} - \sigma_{SB}\epsilon_{em}\big(T^4-T_{env}^4\big)\bigg),
\end{equation}
where $C_v$ is the specific heat, $E_{ion}$ is the nominal ion energy, $\sigma_{SB} \approx 3.54\times 10^{-7}$eV$\cdot$s$^{-1}\cdot$nm$^{-2}\cdot$K$^{-4}$ is the Stefan-Boltzmann constant, $T$ is the wafer temperature, and $T_{env}$ is the temperature of the environment. This leads to a steady-state temperature $T = \big(T_{env}^4 + \frac{fE_{ion}\cos(\theta)}{\sigma_{SB} \epsilon_{em} }\big)^{1/4}$, which we use throughout the present work. We use $\epsilon_{em} =.75$ as an estimate for the effective emissivity of a typical irradiated Si wafer in an open-back wafer holder \cite{parry-JVST-1976}. We take $T_{env} = 294$ K unless otherwise stated. These parameter values lead to reasonable agreement with heating observed during ion bombardment in \cite{madi-thesis-2011,perkinson-JVSTA-2013-Kr-Ge,ishii-etal-JMR-2014,perkinsonthesis2017} and elsewhere and are sufficient for our present purposes.

\paragraph{Estimating translation speed $V(\theta)$.} We require an estimate of the speed of the film's downward (i.e., into-the-substrate) translation due to erosion, $V(\theta)$ \cite{Swenson_2018,evans-norris-JPCM-2022,evans-norris-JPCM-2023,evans-norris-JEM-2024}. Rather than performing a large number of simulations in, e.g., \cite{ziegler-biersack-littmark-1985-SRIM,srim-2000.40}, or other BCA tools, we use the semi-empirical Yamamura approximation of sputter yield $Y(\theta)$ \cite{yamamura-etal-1983-IPP,yamamura-etal-RE-1987},
\begin{equation}
    Y(\theta)= Y(0)\sec^{f_Y}(\theta)\exp(-\Sigma_Y(\sec(\theta)-1)\big),
\end{equation}
which has been commonly used; see, for example, \cite{holmes-cerfon-etal-APL-2012,holmes-cerfon-etal-PRB-2012,pearson-bradley-JPCM-2015}. Above, $f_Y = f_S\big(1 + 2.5\big(\frac{1-\xi}{\xi}\big)\big)$, $\xi = 1 - \sqrt{\frac{E_{th}}{E_{ion}}}$, $\Sigma_Y = f_Y\cos(\theta_{opt})$, $\theta_{opt}=90^{\circ}-286\psi(E_{ion})^{0.45}$, and $\psi(E_{ion})=\frac{\psi(1 \text{ eV})}{\sqrt{E_{ion}}}$. Parameters $E_{th}, f_S$, and $\psi(1 \text{ eV})$ are ion- and target- dependent quantities that can be looked up in the tables of \cite{yamamura-etal-1983-IPP}. For Ar$^+$-irradiated Si, $E_{th}=33$eV, $f_S=1.85$ and $\psi(1\text{ eV}) = 0.0895$ \cite{yamamura-etal-1983-IPP}. To compute the normal-incidence sputter yield $Y(0)$ for any particular $E_{ion}$, we again use the formulas provided in \cite{yamamura-etal-1983-IPP}, which we do not record here for brevity. Then $V(\theta)=f\cos(\theta)Y(\theta)V_0$ is easily computed for Ar$^+$-irradiated Si at any flux, angle, and energy of interest.

\paragraph{Scaling of $a,\alpha,\beta, Y_V$ and $Y_I$ with $E_{ion}$.} Elsewhere \cite{ziberi-etal-PRB-2005}, it has been observed that ion implantation parameters $a,\alpha,\beta$ scale roughly with $\sqrt{E_{ion}}$ for a fixed ion-target combination at low energies, and have negligible angle-dependence \cite{hossain-etal-JAP-2012,hobler-etal-PRB-2016} for irradiation angles not ``too close" to grazing. In the Results section, we make use of this simple scaling argument to obtain results for Ar$^+$-irradiated Si at different energies by scaling $a,\alpha$ and $\beta$ obtained for 1keV Ar$^+$-irradiated Si using SRIM \cite{ziegler-biersack-littmark-1985-SRIM,srim-2000.40}. To scale per-ion defect production rates $Y_V$ and $Y_I$ with energy, we apply the usual Kinchin-Pease relationship \cite{kinchin-pease-1955,liedke-thesis-2011,norris-etal-NCOMM-2011}, $Y_V = \frac{0.8E}{E_D}, Y_I = \frac{0.8E}{E_D}$, where $E_D= 15$eV is the displacement energy for a Si atom.

\paragraph{Numerical implementation.} The nonlinear system of boundary value problems of Equations (\ref{steadystateCI0})-(\ref{SSlowerBCs}) is formidable; we make no serious attempt at solving it using analytical techniques. Instead, we turn to numerical solution using Matlab's \textit{bvp5c}, first numerically solving (\ref{steadystateCI0})-(\ref{SSlowerBCs}) and then passing steady-state solutions to a second solver for the eigenvalue problem represented by the $\mathcal{O}(\epsilon)$ equations discussed in the Appendix. This produces a full set of solutions $w_0(z)$, $u_0(z)$, $p_0(z)$, $\rho_0(z)$, $\eta_0(z)$, $C_{V0}(z)$, $C_{I0}(z)$ in the steady state, and $\tilde{w}_1(z)$, $\tilde{u}_1(z)$, $\tilde{p}_1(z)$, $\tilde{\rho}_1(z)$, $\tilde{\eta}_1(z)$, $\tilde{C}_{V1}(z)$, $\tilde{C}_{I1}(z)$ and $\Sigma$ for the $\mathcal{O}(\epsilon)$ equations. To avoid loss of numerical precision, all stated parameter values are converted to units of \AA, ns, K and Gamu (i.e., $10^{9}$ amu) in Matlab.

\section{Results}
The model (\ref{stresstensor})-(\ref{lowerBCs}) contains too many adjustable parameters for an exhaustive survey of parametric dependence. To reduce the number of free parameters, we model the irradiation of Si by Ar$^+$, where many parameter values have already been estimated. Accordingly, we assign $V_I=V_0=-V_v$, and $V_0=0.02$ nm$^3$ and atomic mass $m_0 =28.09$ amu, corresponding to Si. Since interstitials and vacancies occur in pairs \cite{kinchin-pease-1955}, we assign $Y_V=Y_I$. We also assign $\lambda=.237$ nm, representing a typical nearest-neighbor distance between silicon atoms in a crystalline structure. As in \cite{chan-chason-JVSTA-2008}, we find that the choice of $\lambda$ is not a strong influence on our results.

We assign $\eta^*=10^{5}$ GPa$^{-1}$$\cdot$s$^{-1}$ so that un-irradiated silicon has an extremely high baseline viscosity exceeding experimental estimates by two or more orders of magnitude; see, e.g., \cite{hofsass-bobes-zhang-JAP-2016,perkinsonthesis2017,norris-etal-SREP-2017}. We also assign the fluidity-per-defect parameters $\eta^{-1}_I=\eta^{-1}_V=2.39$ nm$^3\cdot$GPa$^{-1}$$\cdot$s$^{-1}$, as estimated in \cite{ishii-etal-JMR-2014}, where Ar$^+$-irradiated Si was studied. To compute temperature-dependent diffusivities $D_V$ and $D_I$, we apply Arrhenius relations, $D_{V}(T) = D_{V0}e^{-\frac{Q_r}{K_BT}}$ and $D_{I}(T) = D_{I0}e^{-\frac{Q_r}{K_BT}}$ where $T$ is computed as described in the Model section above. We assign $Q_r=2$ eV, within the range of values discussed in, e.g., \cite{fahey-et-al-RoMP-1989,witvrouw-spaepen-JAP-1993b}. We choose plausible coefficients $D_{I0}=10^{25}$nm$^2$/ns and $D_{V0}=10^{27}$nm$^2$/ns so that $D_I \approx 10^{-8}$nm$^2$/ns and $D_V\approx10^{-6}$nm$^2$/ns at room temperature, within the range of values from \cite{bronner-plummer-JAP-1987,fahey-et-al-RoMP-1989,coffa-libertino-APL-1998,ishii-etal-JMR-2014}. Surface energy $\gamma$ is assigned as $1.36 \frac{\text{J}}{\text{m}^2}$ \cite{vauth-mayr-PRB-2007,vauth-mayr-PRB-2008b}. We will also consider $D_{I0}=10^{23}$nm$^2$/ns and $D_{V0}=10^{25}$nm$^2$/ns to better understand the influence of lower diffusivities.

\subsection{Linear stability results}

\begin{figure}
\includegraphics[width=1\linewidth]{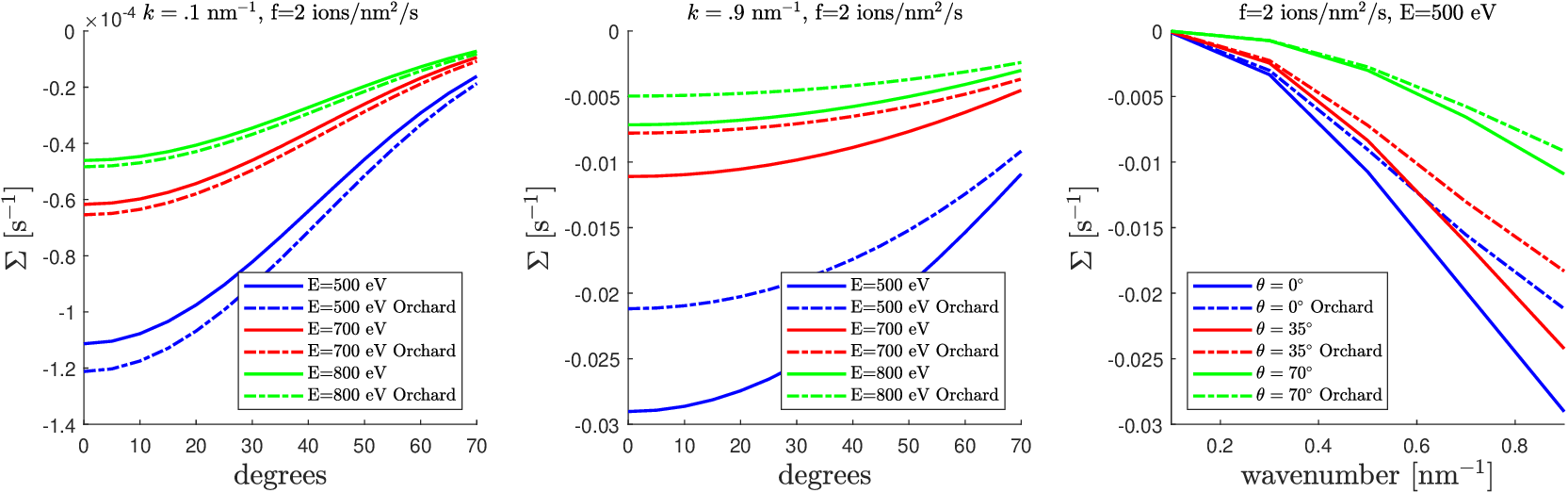}
\caption{Comparison of $\Sigma_{PM}$ and $\Sigma_{Orchard}$ for selected energies, wavenumbers, and irradiation angle. There is broad agreement between the two, which worsens somewhat at higher wavenumbers and lower fluxes. An intuitive explanation for this is discussed in the main text. }
\label{fig:resultsfig1}
\end{figure}

\begin{figure}
\includegraphics[width=1\linewidth]{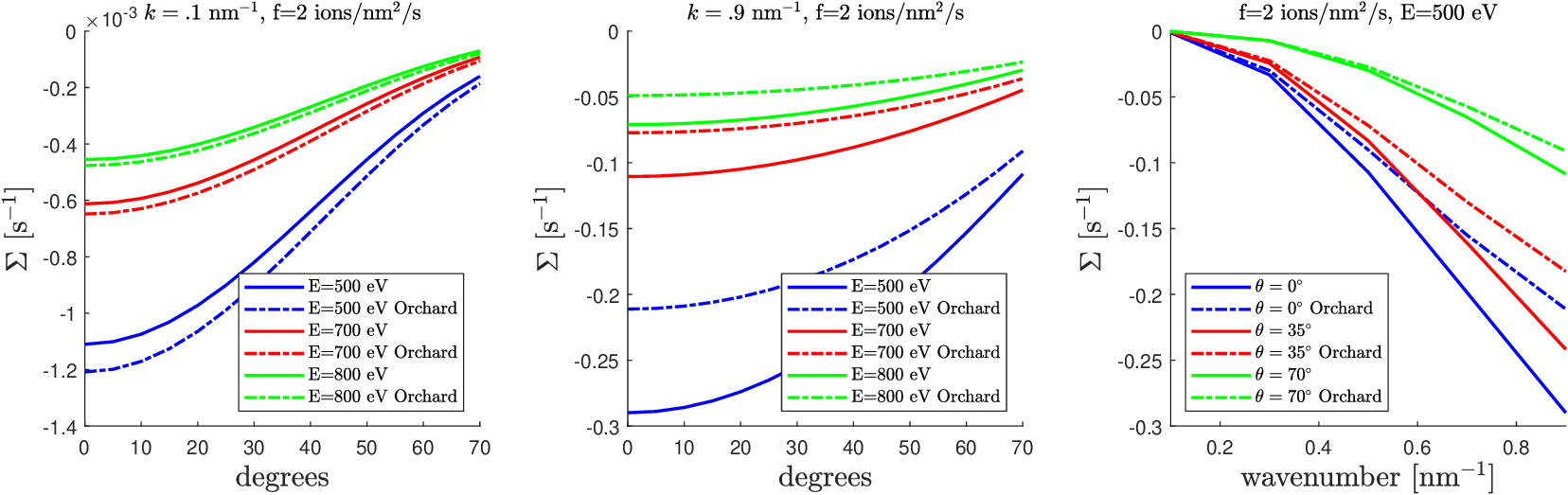}
\caption{Comparison of $\Sigma_{PM}$ and $\Sigma_{Orchard}$, as in Figure \ref{fig:resultsfig1}, while using $D_{A0}$ and $D_{B0}$ two orders of magnitude lower. The curves are indistinguishable up to a scaling factor of $10^{-1}$. }
\label{fig:resultsfig2}
\end{figure}

We compare the numerical results for linear growth rate $\Sigma$ of the present model, which we denote as $\Sigma_{PM}$, with the classical Orchard result \cite{orchard-ASR-1962} for that of a fluid of uniform $\eta^{-1}$,
\begin{equation}
\Sigma_{\text{Orchard}}  = -\frac{\gamma \eta^{-1} }{2 h_{0}(\theta)} \frac{Q(\sinh(2Q) - 2Q)}{1+2Q^2+\cosh(2Q)}\label{orchard}.
\end{equation}
Above, $Q=kh_0(\theta)$ and we assign the uniform fluidity $\eta^{-1}$ as $\langle \eta^{-1} \rangle$, the mean fluidity across the film depth, computed using the present model. This allows for a fair, if somewhat simplified, quantitative comparison between predicted relaxation rates using \cite{orchard-ASR-1962} and those of the present model. In Figure \ref{fig:resultsfig1}, we compare $\Sigma_{\text{Orchard}}$ with $\Sigma$ of the present work for a fixed flux $f$ and a few wavenumbers $k$, ion energies $E$, and angles of incidence $\theta$, with other parameter values assigned as discussed above. In Figure \ref{fig:resultsfig2}, we produce results for the same set of parameters, except with the basic diffusivities $D_{V0}$ and $D_{I0}$ assigned two orders of magnitude lower: $D_{I0}=10^{23}$nm$^2$/ns and $D_{V0}=10^{25}$nm$^2$/ns instead of $D_{I0}=10^{25}$nm$^2$/ns and $D_{V0}=10^{27}$nm$^2$/ns. Here, a few observations can be made:
\begin{itemize}
    \item The numerically-computed $\Sigma_{PM}$ exhibits the same form as that of $\Sigma_{\text{Orchard}}$ up to the assignment of $\eta^{-1}$. At higher energies, the agreement is stronger; we attribute this to thermally-enhanced diffusion tending to uniformize the distribution of defects, leading to a more uniform fluidity.
    \item Differences between $\Sigma_{PM}$ and $\Sigma_{\text{Orchard}}$ are more pronounced at lower energies and at higher wavenumbers. The intuitive explanation for the energy dependence rests on noting that higher energies lead to greater temperatures, which enhance diffusion, and lead to more uniform $\eta^{-1}$ across the film. For the wavenumber dependence, one notes that the effect of geometric flux dilution \cite{NorrisAziz_predictivemodel,evans-norris-arxiv-2025b} on the distribution of deposited energy in the amorphous layer is least for flat surfaces ($k \to 0$), and increases with $k$. However, for most wavenumbers experimentally observed (below around .4 nm$^{-1}$, corresponding to wavenumbers larger than 15nm \cite{castro-cuerno-ASS-2012,castro-etal-PRB-2012,perkinson-JVSTA-2013-Kr-Ge,moreno-barrado-etal-PRB-2015,norris-etal-SREP-2017}), the difference is on the order of 10\%. 
    \item The effect of decreasing diffusivities by two orders of magnitude is simply to reduce $\Sigma_{PM}$ by \textit{one} order of magnitude. This observation is immediately useful --- despite the difficulty surrounding parameter estimation for models of irradiated thin films \cite{NorrisAziz_predictivemodel}, any qualitative results that we obtain here should survive refinements of diffusivity values. However, the present model, with $D_{V0}$ and $D_{I0}$ tuned to produce reasonable agreement with \cite{coffa-libertino-APL-1998} at room temperature, leads to $\Sigma_{PM}$ of the same order of magnitude as experimentally observed in \cite{norris-etal-SREP-2017}. This also lends some confidence that the diffusivities used here are reasonable.
\end{itemize}

Hence the \textit{form} of Equation \ref{orchard} is essentially correct, supposing that the value of $\langle \eta^{-1} \rangle$ is known.

\subsection{Results for the fluidity} 

\begin{figure}
\includegraphics[width=1\linewidth]{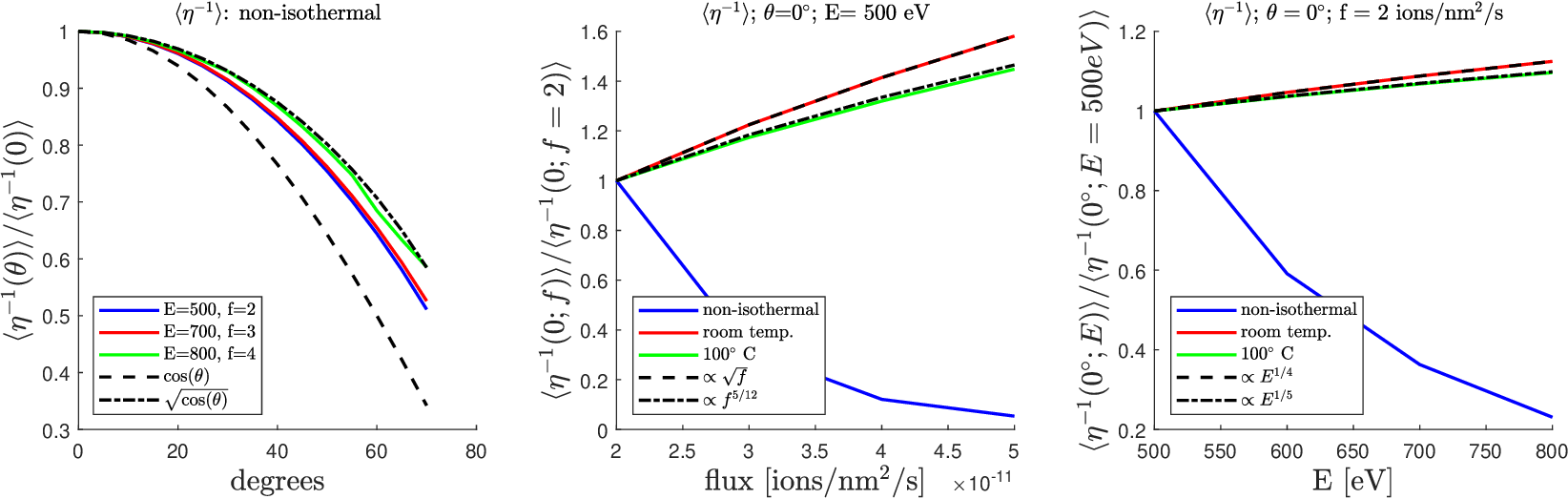}
\caption{\textbf{Left}: angle-dependence of $\eta^{-1}$ for selected energies and fluxes. For comparison, $\cos(\theta)$ and $\sqrt{\cos(\theta)}$ are plotted. The data is evidently consistent with a $\sqrt{\cos(\theta)}$ dependence at these energies and fluxes. \textbf{Center}: flux dependence under selected thermal conditions. When temperature is controlled, and especially near room temperature, $\eta^{-1}$ scales approximately as $\sqrt{f}$. \textbf{Right}: energy dependence of $\eta^{-1}$. When temperature is controlled, and especially near room temperature, $\eta^{-1}$ scales with $E^{1/4}$.}
\label{fig:resultsfig3}
\end{figure}

When modeling irradiated surfaces as viscous thin films, an estimate of $\eta^{-1}$ is necessary to obtain predictions of angle-dependence wavelengths, bifurcation angle, roughening, and other common points of comparison between theory and experiment. Its parametric dependence on $f$, $E$, $\theta$ has been a subject of interest and speculation. Suggestions for flux dependence have included $\eta^{-1}(\theta)/\eta^{-1}(0) \propto f\cos(\theta)$, $\eta^{-1}(\theta)/\eta^{-1}(0)$ constant, and $\eta^{-1}(\theta)/\eta^{-1}(0) \propto \sqrt{f}$. $\eta^{-1}$ has also been broadly assumed to scale with $\frac{E}{h_0}$, which would suggest a roughly $\sqrt{E}$ dependence for constant $f$ and $\theta$.

In Figure \ref{fig:resultsfig3}, we show the angle, flux, and energy dependence of $\langle \eta^{-1} \rangle$ under various experimental conditions. Several aspects of the data are noteworthy.
\begin{itemize}
    \item For the assigned parameters $D_{V0}, D_{I0}, Q_r$, increased flux can lead to \textit{reductions} in fluidity when the increased heating produces much faster annealing of defects. However, when temperature is controlled (here, either to room temperature, or to 100$^{\circ}$ C), clear flux scalings can be identified. In particular, we observe a $\sqrt{f}$ scaling of $\langle \eta^{-1} \rangle$ at fixed energy and irradiation angle, consistent with \cite{ishii-etal-JMR-2014}.
    \item Across fluxes and energies, an angle dependence close to $\sqrt{\cos(\theta)}$ --- rather than $\cos(\theta)$ --- is observed. Taken alongside the above observation about $f$, this suggests a flux-diluted scaling like $\sqrt{f\cos(\theta)}$ for constant energy, regardless of heating.
    \item When temperature is controlled, energy dependence is far weaker than previously believed, showing an $E^{1/4}$ scaling at constant flux and irradiation angle. Otherwise, increased energy leads to increased heating, which can anneal away defects by enhancing diffusion and recombination. This result is again consistent with the fluidity scaling implied by \cite{ishii-etal-JMR-2014}; see also \cite{evans-norris-arxiv-2025}.
\end{itemize}

\section{Discussion}

\paragraph{Conclusions.} In this work, we have developed and studied a model that combines a common modeling simplification --- treating the amorphous layer as a viscous thin film --- with a model for defect kinetics, where the local defect concentration determines local fluidity. Radiation-induced heating was taken into account by adjusting the diffusivity rates of the defects. In an effort to minimize free parameters and obtain physically meaningful results, we have used plausible parameter values, while acknowledging that many are difficult to obtain, and remain uncertain.

We have obtained two sets of results: one pertaining directly to the mean ion-enhanced fluidity, $\langle \eta^{-1}\rangle$, and its angle, flux, and energy dependence in the steady state, and one pertaining to the linear stability effects of viscous surface relaxation in the presence of spatially inhomogeneous defects. Our main findings are summarized as follows.
\begin{itemize}
    \item The form (\ref{orchard}) is consistent with surface relaxation of an amorphous thin film even in the presence of defects, provided that $\eta^{-1}$ can be assigned. In particular, differences between (\ref{orchard}) and the decay rate of surface perturbations predicted by the present work disappear in the limit of small wavenumber $k$. Hence the long-wave approximation of (\ref{orchard}) is validated for seeking theoretical predictions of bifurcation angles.
    \item In contrast with commonly used scaling arguments, we find that depth-averaged fluidity, $\langle \eta^{-1}\rangle$, near room temperature is expected to scale with flux as $\sqrt{f}$, with angle as $\sqrt{\cos(\theta)}$, and with energy as $E^{1/4}$. This is consistent with $\eta^{-1}(\theta,f,E) \propto \frac{\sqrt{f\cos(\theta)}E(\theta)}{h_0(\theta)}$, an intriguing modification of the scaling arguments that have been considered elsewhere \cite{norris-etal-NCOMM-2011,hofsass-APA-2014,hofsass-APA-2015,hofsass-bobes-zhang-JAP-2016,norris-etal-SREP-2017}.
    \item Using plausible values of $Q_r,D_V,D_I$, and $\epsilon_{em}$, thermal effects can lead to a \textit{loss} of fluidity as $f$ and $E$ are increased, even at temperatures typically reached during irradiation. This would be expected to exert a strong influence on pattern formation in typical experimental systems.
\end{itemize}

While the present results are sensitive to the diffusion rates of vacancies and interstitials, and especially to how the diffusion rates respond to temperature changes, these findings are significant for efforts to model ion-induced pattern formation on semiconductor surfaces like silicon and germanium. In particular, we have shown that the defect kinetics within the amorphous thin film cannot, in general, be ignored for the purposes of modeling pattern formation, especially when temperature is not controlled during irradiation.

\paragraph{Future work.} This work illustrates important connections between defect modeling and the idealization of amorphous thin films as fluids, and exposes the potential for sensitivity of irradiation thin film evolution to the details of the underlying defect kinetics --- especially because fluidity can evidently \textit{decrease} with flux and energy if thermally-enhanced recombination of defects is strong enough. This highlights the importance of estimating $Q_r, D_{V0}$ and $D_{I0}$ at energies relevant to low-energy irradiation experiments. Because these are only three free parameters, they could, in principle, be used to fit a large experimental data set across fluxes and energies. These and other implications await further exploration.

In this work, we have neglected two important effects --- sputter removal of defects and ion implantation --- which would be needed to develop a fully quantitative model capable of predicting stresses and fluidities across fluxes, angles, energies, and temperatures, using only a small set of fit parameters. This is a clear direction for future development. Implications for ion-induced pattern formation in the presence of ion-induced stresses, as modeled by an extension of the present work, should also be studied. 

\section*{Acknowledgments}
TPE gratefully acknowledges support from the National Science Foundation through DMS-2136198. EH gratefully acknowledges support from the Department of Mathematics at the University of Utah and the National Science Foundation through DMS-2136198.

\appendix
\section{Linearization details} Here, we record the details of linearization of Equations (\ref{stresstensor})-(\ref{lowerBCs}).
\paragraph{Steady-state equations.}
Conservation of mass and momentum become:
\begin{equation}
\begin{gathered}
\frac{\partial}{\partial z}(\rho_0 w_0) = 0 \\
\frac{\partial}{\partial z}(\eta_0 u_{0,z}) = 0 \\
-p_{0,z} + 2\frac{\partial}{\partial z}(\eta_0 w_{0,z}) = 0.
\end{gathered}
\end{equation}
The steady-state equations for the defect kinetics are:
\begin{equation}
\begin{gathered}
fY_VP_0(z) + D_{V}\frac{\partial^2}{\partial z^2}C_{V,0} + \frac{1}{3}\bigg(\frac{D_{V}V_{V}}{K_BT}\bigg)C_{V,0}p_{0zz} - 4\pi \lambda C_{I,0}C_{V,0}(D_I + D_V) = 0 \\
fY_IP_0(z) + D_{I}\frac{\partial^2}{\partial z^2}C_{I,0} + \frac{1}{3}\bigg(\frac{D_{I}V_{I}}{K_BT}\bigg)C_{I,0}p_{0zz} - 4\pi \lambda C_{I,0}C_{V,0}(D_I + D_V) = 0 
\end{gathered}
\end{equation}


\noindent At lower interface $z=0$:
\begin{equation}
    \begin{gathered}
        u_0 = 0 \\
        w_0 = V(\theta) \\
    \Delta_0 = \frac{\partial \Delta_0}{\partial z} =
    \frac{\partial^2 \Delta_0}{\partial z^2} = 0 \\
    \end{gathered}
\end{equation}

\noindent At free interface $z=h_0(\theta)$:
\begin{equation}
    \begin{gathered}
        0 = w_0 - V(\theta)\frac{\rho^*}{\rho_0} \\
        \eta_0 u_{0,z} = 0 \\
        -p_0 + 2\eta_0w_{0,z} = 0
    \end{gathered}
\end{equation}

\paragraph{Note on the steady-state equations.} Although the steady-state equations appear to be analytically intractable, two useful insights are readily available. 
\begin{itemize}
    \item First, observe that $\rho_0w_0 = c_1$, $\eta_0u_{0,z} = c_2$, $-p_0 + 2\eta_0w_{0,z}=c_3$, and the boundary condition at $z=h_0(\theta)$ implies $c_3=0$. Also, $w_0 = \frac{c_1}{\rho_0}$, implying that $w_{0z} = -\frac{c_1\rho_{0z}}{\rho_0^2}$, and the boundary conditions at $z=0$ imply that $c_1=V(\theta)\rho^*$. Then the steady pressure $p_0= -2V(\theta)\rho^*\frac{\eta_0\rho_{0z}}{\rho_0^2}$, and $\eta_0$ and $\rho_0$ are defined entirely in terms of $C_{V0}$ and $C_{I0}$. Moreover, the steady-state equations for $C_{I0}$ and $C_{V0}$ are independent of the others. Hence the steady state is entirely governed by the defect fields. This reflects the primacy of the defect kinetics in understanding the thin film dynamics.
    \item Second, observe that $u_{0z} = \frac{c_2}{\eta_0}$. Then $u_0 = \int_0^z\frac{c_2}{\eta_0}dz + c_4$. But at $z=0, u_0=0$, so $c_4=0$, and $\eta_0u_{0,z}=0$ at $z=h_0(\theta)$ implies that $c_2=0$. Hence $u_0=0$. Physically, this says that the prescribed defect kinetics cannot produce a shear flow of the type expected from \cite{norris-PRB-2012-linear-viscous} and related work. This observation also permits simplification of the linearized equations.
\end{itemize}

\paragraph{Simplification of steady-state equations with $\frac{D_I}{D_V} \to 0$.} The steady-state equations, as-written, frustrate typical analytical and numerical techniques. In particular, the presence of $p_{0zz}$ in the defect kinetics equations leads to two third-order ordinary differential equations where $C_{I0}'''$ and $C_{V0}'''$ cannot be separated algebraically, as would be required for a numerical solution using $bvp5c$ or other pre-packaged numerical solvers. However, in silicon, self-diffusion of interstitials is orders of magnitude slower than that of vacancies; that is, $\frac{D_I}{D_V} \approx 0$ \cite{coffa-libertino-APL-1998,fahey-et-al-RoMP-1989}. Then we identify a characteristic rate $\frac{D_Vh_0}{V_V}$ (with units $s^{-1}$). Dividing both sides by this, we obtain the dimensionless equations
\begin{equation}
\begin{gathered}
    \frac{fY_V P_0(z)V_V}{D_Vh_0} + \frac{V_V}{h_0}C_{V,0zz} + \frac{1}{3}\frac{V_V}{h_0}\bigg(\frac{V_V}{K_BT}\bigg)C_{V,0}p_{0zz} - 4\pi\lambda C_{V0}C_{I0}\frac{V_V}{h_0}\bigg(1+ \frac{D_I}{D_V}\bigg) = 0 \\
    \frac{fY_I P_0(z)V_V}{D_Vh_0} + \bigg(\frac{D_I}{D_V}\bigg)\frac{V_V}{h_0}C_{I,0zz} + \frac{1}{3}\frac{D_I}{D_V}\frac{V_V}{h_0}\bigg(\frac{V_I}{K_BT}\bigg)C_{I,0}p_{0zz} - 4\pi\lambda C_{V0}C_{I0}\frac{V_V}{h_0}\bigg(1+ \frac{D_I}{D_V}\bigg) = 0.
\end{gathered}
\end{equation}
To leading order in $\frac{D_I}{D_V}$, we obtain
\begin{equation}
    \begin{gathered}
    \frac{fY_V P_0(z)V_V}{D_Vh_0} + \frac{V_V}{h_0}C_{V,0zz} + \frac{1}{3}\frac{V_V}{h_0}\bigg(\frac{V_V}{K_BT}\bigg)C_{V,0}p_{0zz} - 4\pi\lambda C_{V0}C_{I0}\frac{V_V}{h_0} = 0 \\
    \frac{fY_I P_0(z)V_V}{D_Vh_0} - 4\pi\lambda C_{V0}C_{I0}\frac{V_V}{h_0} = 0, \label{ss-reduced}
    \end{gathered}
\end{equation}
with the latter leading immediately to
\begin{equation}
    C_{I0} = \frac{fY_IP_0(z)}{4\pi\lambda C_{V0}D_V} \label{CV0eqn}
\end{equation}
to leading order in the diffusivity ratio. This approximation greatly simplifies the problem at hand. With these simplifications, we express $p_{0zz}$ in terms of $C_{V0}$, and Equation \ref{ss-reduced}a yields a nonlinear third-order ODE, and the three boundary conditions for $\Delta_0$ at $z=0$, when expressed in terms of $C_{V0}$, lead to a solvable problem whose solution fully determines the other steady-state quantities, $u_0,w_0,\rho_0,\eta_0,p_0,C_{V0}$.

\paragraph{Linearized equations with $\frac{D_V}{D_I} \to 0$.} In the bulk, conservation of mass and momentum are
\begin{equation}
\begin{gathered}
\rho_{1t} + \rho_0u_{1x} + \frac{\partial}{\partial z}(\rho_0w_1) + \frac{\partial}{\partial z}(\rho_1w_0) = 0; \\
-p_{1x} + 2\eta_0u_{1xx} + \frac{\partial}{\partial z}\big(\eta_0(u_{1z} + w_{1x})\big) = 0; \\
-p_{1z} + \eta_0(w_{1xx} + u_{1xz}) + 2\frac{\partial}{\partial z}(\eta_0w_{1z}) + 2\frac{\partial}{\partial z}(\eta_1w_{0z}) = 0,
\end{gathered}
\end{equation}
and the reaction-diffusion equations for the defects are
\begin{equation}
\begin{gathered}
C_{V,1t} = fY_VP_1(x,z) + D_{V}\nabla^2C_{V,1} - \frac{1}{3}\bigg(\frac{D_{V}V_{V}}{K_BT}\big(C_{V,1}\nabla^2(-2p_0 + 2\eta_0w_{0z}) \\ + C_{V,0}\nabla^2(-2p_1 + 2\eta_0(u_{1x} + w_{1z}) +2\eta_1(w_{0z})) \bigg) 
-4\pi \lambda D_V(C_{V0}C_{I1} + C_{V1}C_{I0}); \vspace{.5cm} \\
C_{I,1t} = fY_IP_1(x,z) -4\pi \lambda D_V(C_{V0}C_{I1} + C_{V1}C_{I0}).
\end{gathered}
\end{equation}
\noindent At lower interface $z=0$:
\begin{equation}
\begin{gathered}
u_1 = 0 \\
w_{0z}g_1 + w_1 = 0 \\
\Delta_{0z}g_1 + \Delta_1 = 0 \\
\Delta_{0zz}g_1 + \Delta_{1z} = 0 \\
\Delta_{0zzz}g_1 + \Delta_{1zz} = 0
\end{gathered}
\end{equation}

\noindent At free interface $z=h_0$:
\begin{equation}
\begin{gathered}
h_{1t} = w_1 + h_1w_{0z} + V(\theta)\frac{\rho^*}{\rho_0^2}(\rho_{0z}h_1 + \rho_1) \\
h_{1x}p_0 + \eta_0(u_{1z} + w_{1x}) = 0 \\
h_1\frac{\partial}{\partial z}(-p_0 + 2\eta_0 w_{0z}) + -p_1 + 2(\eta_0 w_{1z} + \eta_1w_{0z}) = -\gamma h_{1xx}
\end{gathered}
\end{equation}
Throughout the above,
\begin{equation}
\begin{gathered}
\rho_0 = \frac{m_0}{V_0 + C_{I,0}V_{I} + C_{V,0}V_{V}}, \hspace{.25cm}
\rho_1 = -\frac{m_0(C_{I,1}V_{I} + C_{V,1}V_{V})}{(V_0 + C_{I,0}V_{I} + C_{V,0}V_{V})^2}, \\
\eta_0 = \frac{1}{\frac{1}{\eta^*} + \frac{C_{I,0}}{\eta_I } + \frac{C_{V,0}}{\eta_V } }, \hspace{.25cm}
\eta_1 = - \frac{ \frac{C_{I,1}}{\eta_I } + \frac{C_{V,1}}{\eta_V }}{ (\frac{1}{\eta^*} + \frac{C_{I,0}}{\eta_I } + \frac{C_{V,0}}{\eta_V })^2},
\end{gathered}
\end{equation}

\paragraph{Linearized equations for normal modes with $\frac{D_V}{D_I} \to 0$.} Following the application of the linear stability ansatz to Equations (\ref{stresstensor})-(\ref{lowerBCs}), we obtain the following set of boundary value problems for eigenvalue $\sigma$, which will determine the stability of Fourier modes of wavenumber $k$. In the bulk, we obtain the following ODEs for mass conservation and momentum conservation, respectively:
\begin{equation}
\begin{gathered}
\sigma \tilde{\rho}_1 + ik\rho_0\tilde{u}_1 + \rho_{0z}\tilde{w}_1 + \tilde{w}_{1z}\rho_0 + \tilde{\rho}_{1z}w_0 + w_{0z}\tilde{\rho}_1 = 0; \\
-ik \tilde{p}_1 - 2k^2\eta_0\tilde{u}_1 + \eta_{0z}\big(\tilde{u}_{1z} + ik\tilde{w}_1\big) + \eta_{0}\big(\tilde{u}_{1zz} + ik\tilde{w}_{1z}\big) = 0; \\
-\tilde{p}_{1z} + \eta_0\big(-k^2 \tilde{w}_1 + ik\tilde{u}_{1z}\big) + 2\big(\eta_{0z}\tilde{w}_{1z} + \eta_0\tilde{w}_{1zz}\big) + 2\big(\tilde{\eta}_{1z}w_{0z} + \tilde{\eta}_{1}w_{0zz} \big) = 0. \label{chunk1}
\end{gathered}
\end{equation}
We also obtain the following ODEs for the reaction, advection, and diffusion of defects within the amorphous bulk:
\begin{equation}
\begin{gathered}
\sigma \tilde{C}_{V,1} = fY_V\tilde{P}_1 + D_V\big( -k^2 \tilde{C}_{V,1} + \tilde{C}_{V,1zz}\big) - \frac{1}{3}\frac{D_V V_V}{K_BT}\bigg[-\tilde{C}_{V,1}p_{0zz} \\ + C_{V,0}\bigg(-2\big(-k^2 \tilde{p}_1 + \tilde{p}_{1zz}\big) + 2\big( (ik\tilde{u}_1 + \tilde{w}_{1z})\eta_{0zz} + \eta_0 (-ik^3\tilde{u}_1 + ik\tilde{u}_{1zz} -k^2\tilde{w}_{1z} + \tilde{w}_{1zzz}) \\ + 2\eta_{0z}(ik \tilde{u}_{1z} + \tilde{w}_{1zz}) \big) + 2\big(w_{0z}(-k^2 \tilde{\eta}_1 + \tilde{\eta}_{1zz}) + \tilde{\eta}_1 w_{0zzz} + \tilde{\eta}_{1z}w_{0zz}     \big) \bigg)\bigg] \\ -4\pi \lambda D_V(C_{V,0}\tilde{C}_{I,1} + \tilde{C}_{V,1}C_{I,0}) \label{chunk2}
\end{gathered}
\end{equation}
for the interstitials, and
\begin{equation}
\begin{gathered}
\sigma \tilde{C}_{I,1} = fY_I\tilde{P}_1 -4\pi \lambda D_V(C_{V,0}\tilde{C}_{I,1} + \tilde{C}_{V,1}C_{I,0}) \label{CV1eqn}
\end{gathered}
\end{equation}
for the vacancies. These equations for the interior are accompanied by the following boundary conditions. At $z=0$:
\begin{equation}
\begin{gathered}
\tilde{u}_1 = 0 \\
w_{0z}\tilde{g}_1 + \tilde{w}_1 = 0 \\
\Delta_{0z}\tilde{g}_1 + \tilde{\Delta}_1 = 0 \\
\Delta_{0zz}\tilde{g}_1 + \tilde{\Delta}_{1z} = 0 \\
\Delta_{0zzz}\tilde{g}_1 + \tilde{\Delta}_{1zz} = 0.
\end{gathered}
\end{equation}
To facilitate numerical solution, we note that the latter three boundary conditions above can also be expressed as 
\begin{equation}
\begin{gathered}
\tilde{C}_{V1} = -\bigg[C_{V0z} + C_{I0z}\frac{V_I}{V_V}\bigg]\tilde{g}_1 - \tilde{C}_{I1}\frac{V_I}{V_V} \\
\tilde{C}_{V1z} = -\bigg[C_{V0zz} + C_{I0zz}\frac{V_I}{V_V}\bigg]\tilde{g}_1 - \tilde{C}_{I1z}\frac{V_I}{V_V} \\
\tilde{C}_{V1zz} = -\bigg[C_{V0zzz} + C_{I0zzz}\frac{V_I}{V_V}\bigg]\tilde{g}_1 - \tilde{C}_{I1zz}\frac{V_I}{V_V}.
\end{gathered}
\end{equation}
At $z=h_0$:
\begin{equation}
\begin{gathered}
\sigma \tilde{h}_1 = \tilde{w}_1 + \tilde{h}_1w_{0z} + V(\theta)\frac{\rho^*}{\rho_0^2}(\rho_{0z}\tilde{h}_1 + \tilde{\rho}_1  )  \\
ik \tilde{h}_1p_0 + \eta_0(\tilde{u}_{1z} + ik\tilde{w}_1) = 0  \\
\tilde{h}_1\big(-p_{0z} + 2(\eta_{0z}w_{0z} + \eta_0 w_{0zz})\big) -\tilde{p}_1 + 2(\eta_0\tilde{w}_{1z} + \tilde{\eta}_1w_{0z}) = \gamma k^2\tilde{h}_{1}.
\end{gathered}
\end{equation}
Throughout the above,
\begin{equation}
\begin{gathered}
\tilde{\rho}_1 = -\frac{m_0(\tilde{C}_{I,1}V_{I} + C_{V,1}V_{V})}{(V_0 + C_{I,0}V_{I} + C_{V,0}V_{V})^2}, \hspace{.25cm}
\tilde{\eta}_1 = - \frac{ \frac{\tilde{C}_{I,1}}{\eta_I } + \frac{\tilde{C}_{V,1}}{\eta_V }     }{ (\frac{1}{\eta^*} + \frac{C_{I,0}}{\eta_I } + \frac{C_{V,0}}{\eta_V })^2 },
\end{gathered}
\end{equation}
where we again note that $C_{V0}$ and $C_{V1}$ are fully determined by $C_{I0}$ and $C_{I1}$ according to Equations (\ref{CV0eqn}) and (\ref{CV1eqn}), respectively. We also apply the phase-shift that relates the interfaces,
\begin{equation}
    \tilde{g}_1 = \tilde{h}_1\exp(-ikx_0(\theta)),
\end{equation}
where $x_0(\theta)$ is as described in Equation (\ref{interfaces}). This assignment is described in more detail in \cite{evans-norris-JEM-2024}.

\paragraph{Note on solution technique.} Solving a nonlinear eigenvalue problem with Matlab's bvp5c requires that there be one more boundary condition than the number of ODEs when all of the bulk equations are expressed as a system of first-order ODEs. The presence of eight boundary conditions then requires the expression of the equations for the bulk as a system of seven first-order ODEs. To obtain these, we
\begin{itemize}
    \item use Equation \ref{chunk1}a to express $\tilde{w}_{1z}$ in terms of $\tilde{\rho}_{1z}$ and $\tilde{u}_1$. $\tilde{\rho}_1$ can be expressed in terms of $\tilde{C}_I$ and $\tilde{C}_V$;
    \item use Equation \ref{chunk1}b as an equation for $\tilde{u}_{1zz}$;
    \item use Equation \ref{chunk1}c as an equation for $\tilde{p}_{1z}$, where the $\tilde{w}_{1zz}$ is expressible in terms of $\tilde{C}_V,\tilde{C}_{Vz},\tilde{C}_{Vzz}$, $\tilde{u}_1$ and $\tilde{u}_{1z}$ using Equation \ref{chunk1}a, as described above;
    \item use Equation \ref{chunk2} as a third-order ODE for $\tilde{C}_V$, where $\tilde{p}_{1zz}$, $\tilde{w}_{1zzz}$, $\tilde{u}_{1z}, \tilde{u}_{1zz}$, $\tilde{\eta}_{1z}$ and $\tilde{\eta}_{1zz}$ can be expressed in terms of lower-order derivatives using the other substitutions described above. 
\end{itemize}
In the resulting system of ODEs, each unknown function's highest derivative occurs exactly once, and the resulting equations are expressible as a system of first-order ODEs, as \textit{bvp5c} requires. The manipulations described above were performed using Matlab's Symbolic Math Toolbox. We do not record the resulting equations which were used with bvp5c due to their extreme length ($>$ 50 pages when copied and pasted into a word processor).


\bibliographystyle{plain}
\bibliography{bapaperrefs3}

\end{document}